# Trap-Enhanced Steep-Slope Negative-Capacitance FETs Using Amorphous Oxide Semiconductors

*Yungyeong Park[1], Hakseon Lee[1], and Yeonghun Lee[1,2,3]**

[1]Department of Electronics Engineering, Incheon National University, Incheon 22012, Republic of Korea

[2]Department of Intelligent Semiconductor Engineering, Incheon National University, Incheon 22012, Republic of Korea

[3]Research Institute for Engineering and Technology, Incheon National University, Incheon 22012, Republic of Korea

*Corresponding author. E-mail: y.lee@inu.ac.kr

## ABSTRACT

Amorphous oxide semiconductors (AOSs) have recently gained attention as a promising channel material of back-end-of-line (BEOL)-compatible transistors for monolithic three-dimensional (3D) integrations. However, the degradation in device performance resulting from the high trap densities in AOS, compared to conventional crystalline channel materials, has remained an intractable issue. We introduce the negative-capacitance (NC) operation into the AOS-based transistors. Negative-capacitance field-effect transistors (NCFETs) have been proposed for low-power devices, enabling sub-60 mV/decade subthreshold swing $SS$ induced by a ferroelectric layer. In this work, we develop an AOS NCFET model to investigate the influence of traps within the channel on the steep-slope operation. It is revealed that as the trap density of the channel increases, $SS$ of the MOSFET increases, while the $SS$ of the NCFET decreases. The physical interpretation for steep $SS$ is attributed to the fact that the trapped charges enhance the negative potential drop of the NC layer, enabling the abrupt device switching. This finding will accelerate the development of BEOL transistors and other applications based on the AOS materials in conjunction with the NC effect.

## 1. INTRODUCTION

Since the initial demonstration of amorphous indium-gallium-zinc oxide (a-IGZO) in 2004,[1] numerous studies have been conducted on amorphous oxide semiconductors (AOSs) for flat panel display applications.[2,3] It is due to their superior properties in comparison to the hydrogenated amorphous silicon (a-Si:H), such as enhanced stability, good uniformity, high mobility, low off-current, and high optical transparency.[4–6] AOSs with these advantages have been applied as a channel for thin-film transistors (TFTs) in active-matrix (AM) display, including organic light-emitting diode (OLED) display and liquid crystal display (LCD).[7–9] Presently, the application range of AOSs has extended to back-end-of-line (BEOL)-compatible electronics. The geometrical scaling down of complementary metal-oxide-semiconductor (CMOS) technology to meet the demand of Moore's Law has reached fundamental limits with regard to leakage current and power consumption.[10] In this context, it is highly desirable to implement a computer-in-memory (CIM) architecture through a monolithic 3D integration (M3D), which incorporates logic circuits and memory devices using BEOL-compatible transistors.[11,12] However, the device layers formed after the CMOS fabrication process are vulnerable to thermal degradation, which gives rise to significant limitations in channel materials that satisfy the tight thermal budget (< 400 °C) for further processes integrating BEOL transistors.[13] Nevertheless, the outstanding properties of the IGZO channel with low-temperature processability,[14] high mobility,[15] and low leakage current (by wide band gap)[16] enable the vertical integration of IGZO with the CMOS technology, including processors,[17] image sensors[18] and non-volatile memory (NVM) devices such as ferroelectric RAM (FeRAM),[19] resistive RAM (ReRAM),[20] and 3D NAND[21] with high performance. Furthermore, it has been recently demonstrated that IGZO retains excellent material properties such as a high on/off ratio and high mobility even at extremely small channel length and thickness, which enables the fabrication of highly scaled IGZO transistors for advanced semiconductor applications.[22,23]

Despite the numerous advantages of material properties in IGZO, the device performance is degraded due to the presence of significant traps in the channel.[24] To overcome this challenge, we introduced a negative capacitance in the IGZO transistors by adding a ferroelectric layer in the gate stack. Negative capacitance field effect transistors (NCFETs) are promising candidates for low-power applications due to their ability to switch rapidly at low supply voltage (VDD).[25–27] The polarization response of the ferroelectric material to an external electric field induces the negative capacitance effect, which amplifies the gate voltage and achieves a sub-60 mV/dec swing.[28] To utilize NCFETs efficiently, design conditions for sub-60 $SS$ and, more importantly, hysteresis-free operation should be satisfied through precise capacitance matching.[29] For the capacitance matching, it is essential to secure sufficient semiconductor capacitance for reducing the $SS$ below the Boltzmann tyranny. However, state-of-the-art FET structures with an ultra-thin body (UTB), such as double gate (DG) MOSFETs, FinFETs, and 2D material-based FETs, have minimal parasitic capacitance compared to the gate capacitance due to low channel doping and good electrostatic characteristics, limiting $SS$ to the 60 mV/dec for NCFETs. Meanwhile, it has been proposed that the interface trap between the channel surface and the oxide of NCFETs provides adequate capacitance, thereby enabling steep slope operation.[30,31] This suggests a need to investigate the AOS NCFET, as it is predicted that traps within the channel will play a critical role in improving the $SS$ along with the NC effect.

In this work, we present AOS NCFETs with fast switching and low-voltage operation that benefit from traps and NC effects, thereby resolving the $SS$ limitation of existing transistors. First, we developed a compact model of the NCFETs, considering the traps. We then compared the $SS$ characteristics of the AOS NCFETs and MOSFETs with different trap densities. By increasing the trap density, $SS$ is reduced below 60 mV/dec in the AOS NCFETs. Furthermore, we elucidated the steep slope mechanism by investigating how the trapped charge behaves differently in the gate stack with and without a ferroelectric. Finally, we concluded AOS NCFETs' excellent performance provides potential avenues for the application of BEOL-compatible devices.

## 2. MODELING SECTION

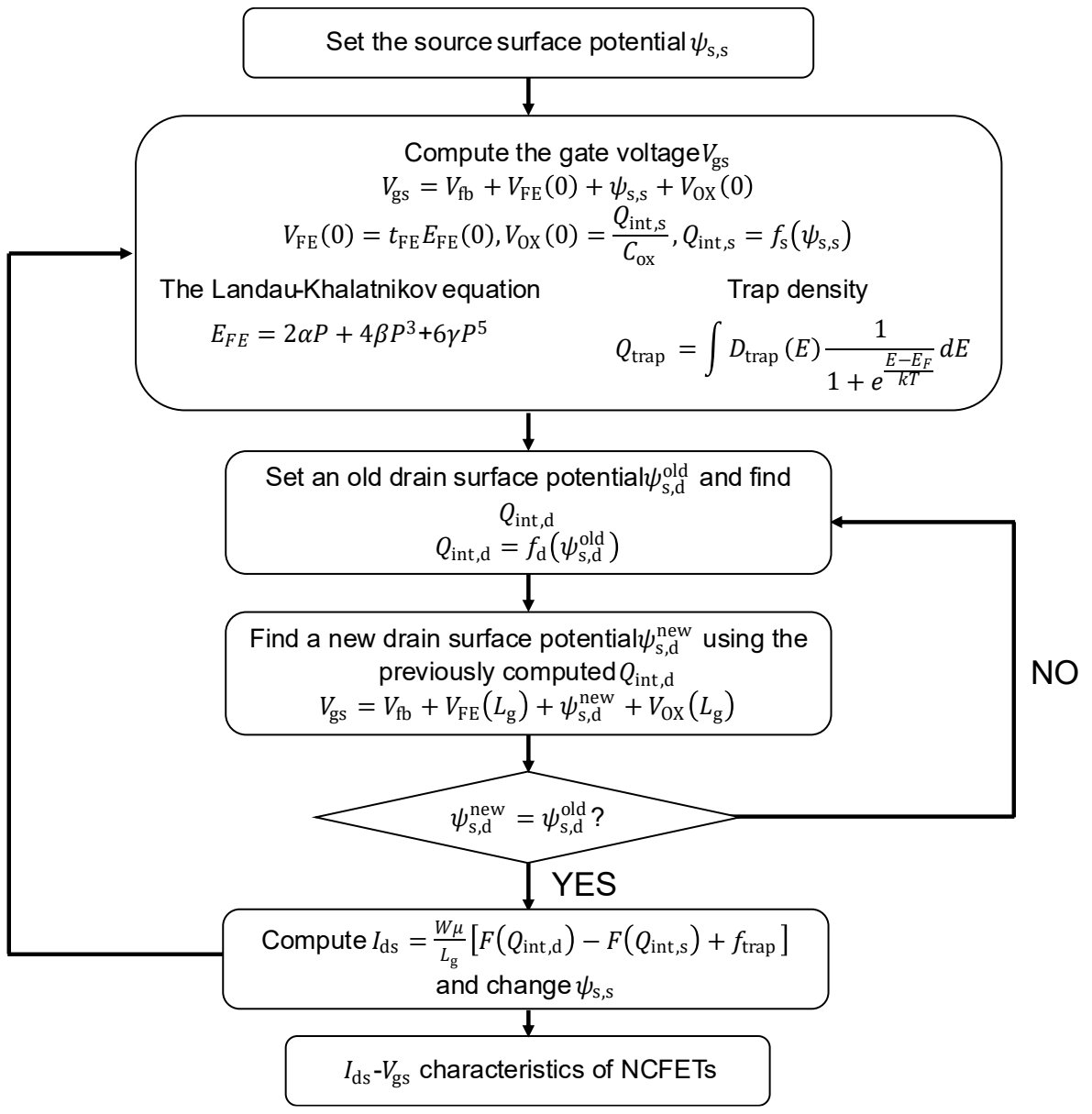

Figure 1. Modeling flow chart for obtaining $I_{ds}$-$V_{gs}$ characteristics of NCFETs including traps.

We developed the 2-D AOS NCFETs model with the metal-ferroelectric-insulator-semiconductor (MFIS) device structure based on the Landau-Khalatnikov (L-K) equation,[32] according to the modeling proposed in [33]. Figure 1 depicts the overall simulation procedure of the model. The L-K equation, which determines the relationship between the electric field $E$ and the polarization $P$ in a ferroelectric material, is given by

$$E = 2\alpha P + 4\beta P^3 + 4\beta P^3 + 6\gamma P^5, \tag{1}$$

where $\alpha$, $\beta$ and $\gamma$ are the ferroelectric material parameters. In this paper, the coefficients of HZO are set to $\alpha = -1.28 \times 10^{10}$ $\mathrm{mF^{-1}}$, $\beta = -2.53 \times 10^{12}$ $\mathrm{m^5F^{-1}C^{-2}}$ and $\gamma = 0$ $\mathrm{m^9F^{-1}C^{-4}}$.[34]

Eq 1 derives the voltage drop in ferroelectrics as follows:

$$V_{FE} = T_{FE}E_{FE} = 2\alpha T_{FE}P + 4\beta T_{FE}P^3, \tag{2}$$

The polarization is approximated by the charge ($Q_{int}\varepsilon_0 E_{FE} + P_{FE} \approx P_{FE}$) because $\varepsilon_0 E_{FE}$ is typically negligible compared to $P_{FE}$. Including $V_{FE}$ expressed as a function of the internal charge in the gate stack $Q_{\mathrm{int}}$, the voltage balance in the equivalent circuit shown in Figure 2a leads to the division of the gate voltage $V_{\mathrm{gs}}$ as described in

$$V_{\mathrm{gs}} = V_{\mathrm{fb}} + V_{\mathrm{FE}}(y) + V_{\mathrm{OX}}(y) + \psi_{\mathrm{s}}(y), \quad (3)$$

where $V_{\mathrm{fb}}$ is the flat band voltage, $V_{\mathrm{OX}}(y) = Q_{\mathrm{int}}(y)/C_{\mathrm{OX}}$ is the voltage drop across the oxide, and $\psi_{\mathrm{s}}(y)$ is the surface potential. $Q_{\mathrm{int}}$ is expressed as a function of the position $y$ in the longitudinal channel direction by the distributed charge effect,[35] and $Q_{\mathrm{int}}(0)$ and $Q_{\mathrm{int}}(L_{\mathrm{g}})$ are internal charges at the source end and the drain end, respectively. The $Q_{\mathrm{int}}$ includes the inversion charge $Q_{\mathrm{inv}}$, the depletion charge $Q_{\mathrm{dep}}(= qN_{\mathrm{ch}})$, and the trap charge in channel $Q_{\mathrm{trap}}$:

$$Q_{\mathrm{int}} = -\left(Q_{\mathrm{inv}}(y) + Q_{\mathrm{dep}} + Q_{\mathrm{trap}}(y)\right), \quad (4)$$

where $Q_{\mathrm{inv}}$ and $Q_{\mathrm{trap}}$ can be numerically calculated by integrating the density of states (DOS) over the energy according to the Fermi-Dirac distribution, as shown in Figure 2b. To obtain $Q_{\mathrm{inv}}$, an effective 2-D DOS for the ultrathin channel is employed, with integration ranging from the conduction band edge $E_{\mathrm{C}}$ to $E_{\mathrm{C}} + 6kT$, where $k$ and $T$ are the Boltzmann constant and the room temperature, respectively. For $Q_{\mathrm{trap}}$, we considered acceptor-like traps with a constant trap density of states $D_{\mathrm{trap}}$ energetically distributed from the intrinsic Fermi level $E_{\mathrm{Fi}}$ to $E_{\mathrm{C}}$. This is because electron trapping and detrapping within the energy range near the conduction band edge influence $SS$ in n-channel devices, while deep donor-like traps do not play a role in determining $SS$ as they are occupied during the device operation condition.[36]

The drift-diffusion current equation in [33] is obtained as follows:

$$I_{\mathrm{ds}} = \frac{W}{L_{\mathrm{g}}}\left(D\frac{dQ_{\mathrm{inv}}}{dy} - \mu Q_{\mathrm{inv}}\frac{d\psi_{\mathrm{s}}}{dy}\right) = \frac{W}{L_{\mathrm{g}}}\left(D\frac{dQ_{\mathrm{inv}}}{dQ_{\mathrm{int}}}\frac{dQ_{\mathrm{int}}}{dy} - \mu Q_{\mathrm{inv}}\frac{d\psi_{\mathrm{s}}}{dQ_{\mathrm{int}}}\frac{dQ_{\mathrm{int}}}{dy}\right), \quad (5)$$

where $D = \mu(kT/q)$ is the diffusion coefficient and $\mu$ is the mobility. Substituting $Q_{\mathrm{inv}} = -Q_{\mathrm{int}} - Q_{\mathrm{dep}} - Q_{\mathrm{trap}}(Q_{\mathrm{int}})$ and $\varphi_{\mathrm{s}} = V_{\mathrm{gs}} - V_{\mathrm{fb}} - V_{\mathrm{FE}}(Q_{\mathrm{int}}) - V_{\mathrm{OX}}(Q_{\mathrm{int}})$ into the Eq 5, the drain equation is reformulated as:

$$I_{\mathrm{ds}} = \frac{W\mu}{L_{\mathrm{g}}} \int_{Q_{\mathrm{int,s}}}^{Q_{\mathrm{int,d}}} \left(1 + \frac{dQ_{\mathrm{trap}}}{dQ_{\mathrm{int}}}\right) + \left(Q_{\mathrm{int}} + Q_{\mathrm{dep}} + Q_{\mathrm{trap}}\right)\left(2\alpha T_{\mathrm{FE}} + Q_{\mathrm{int}}^2 + \frac{1}{C_{\mathrm{OX}}}\right) dQ_{\mathrm{int}}. \tag{6}$$

Carrying out the integration with respect to the $dQ_{\mathrm{int}}$, the drain current $I_{\mathrm{ds}}$ with traps is obtained as follows:

$$I_{\mathrm{ds}} = \frac{W\mu}{L_{\mathrm{g}}} \left[F\left(Q_{\mathrm{int,d}}\right) - F\left(Q_{\mathrm{int,s}}\right) + f_{\mathrm{trap}}\right], \tag{7}$$

where

$$F(Q_{\mathrm{int}}) = -3\beta T_{\mathrm{FE}} Q_{\mathrm{int}}^4 + 4\beta T_{\mathrm{FE}} q N_{\mathrm{ch}} Q_{\mathrm{int}}^3 + \frac{1}{2}\left(2\alpha T_{\mathrm{FE}} + \frac{1}{C_{\mathrm{OX}}}\right) Q_{\mathrm{int}}^2 + \left(2\alpha T_{\mathrm{FE}} + \frac{1}{C_{\mathrm{OX}}}\right) q N_{\mathrm{ch}} - \frac{kT}{q} Q_{\mathrm{int}}, \tag{8}$$

$$f_{\mathrm{trap}} = \int_{Q_{\mathrm{int,s}}}^{Q_{\mathrm{int,d}}} 12\beta T_{\mathrm{FE}} Q_{\mathrm{trap}} Q_{\mathrm{int}}^2 + \left(2\alpha T_{\mathrm{FE}} + \frac{1}{C_{\mathrm{OX}}}\right) Q_{\mathrm{trap}} dQ_{\mathrm{int}} + \frac{kT}{q}\left(Q_{\mathrm{trap,d}} - Q_{\mathrm{trap,s}}\right). \tag{9}$$

Here, $L_{\mathrm{g}}$ is the gate length, $W$ is the channel width, and $q$ is the electron elementary charge. $Q_{\mathrm{int,s}}$ ($Q_{\mathrm{int,d}}$) and $Q_{\mathrm{trap,s}}$ ($Q_{\mathrm{trap,d}}$) are the internal and trap charges at the source end and drain ends, respectively. While $F\left(Q_{\mathrm{int,d}}\right)$ and $F\left(Q_{\mathrm{int,s}}\right)$ are derived through analytical integration with respect to the $dQ_{\mathrm{int}}$ over the range from $Q_{\mathrm{int,s}}$ to $Q_{\mathrm{int,d}}$, $f_{\mathrm{trap}}$ is numerically integrated because $Q_{\mathrm{trap}}$ is the function of $Q_{\mathrm{int}}$. The $Q_{\mathrm{trap}}$ is approximated to the $Q_{\mathrm{trap}} = (Q_{\mathrm{trap,s}} + Q_{\mathrm{trap,d}})/(Q_{\mathrm{int}} - Q_{\mathrm{int,d}}) + Q_{\mathrm{trap,d}}$.

To compute the $I_{\mathrm{ds}}$-$V_{\mathrm{gs}}$ characteristics of the NCFETs, we first set the gate bias according to the source surface potential $\psi_{\mathrm{s,s}}$. Since $Q_{\mathrm{int}}(y)$ is a function of $\psi_{\mathrm{s}}$, we also obtain $Q_{\mathrm{int,d}}$ by setting an old drain surface potential $\psi_{\mathrm{s,d}}^{\mathrm{old}}$. Then, a new drain surface potential $\psi_{\mathrm{s,d}}^{\mathrm{new}}$ is calculated by coupling the previously computed $Q_{\mathrm{int,d}}$ and gate bias. $\psi_{\mathrm{s,d}}$ is self-consistently determined until the old and the new values converge. Finally, by putting the computed $Q_{\mathrm{int}}(y)$ under the given bias conditions into $F(Q_{\mathrm{int,d}})$, $F(Q_{\mathrm{int,s}})$, and $f_{\mathrm{trap}}$, the $I_{\mathrm{ds}}$-$V_{\mathrm{gs}}$ characteristic is obtained.

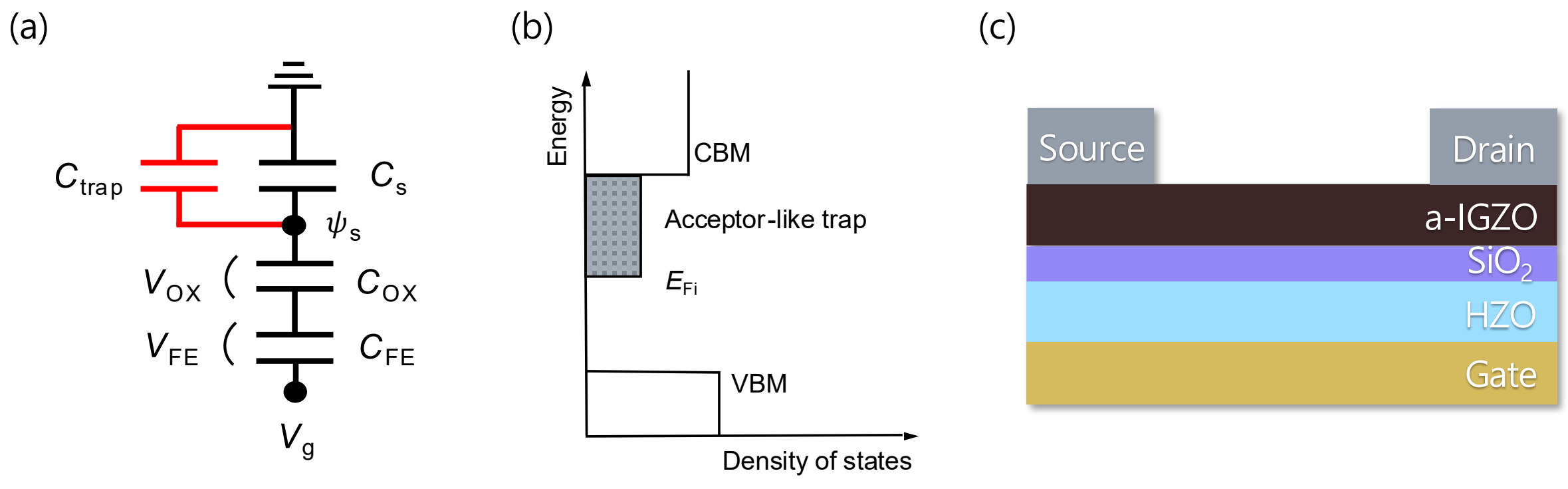

Figure 2. (a) Equivalent capacitor divider circuit model of an NCFET, where $C_{\mathrm{FE}}$, $C_{\mathrm{ox}}$, $C_{\mathrm{trap}}$, and $C_{\mathrm{S}}$ are ferroelectric, oxide, trap, and semiconductor capacitances, respectively. (b) The density of states distribution of the acceptor-like bulk traps. (c) Device structure of the back-gated 2D NCFET with an a-IGZO channel used in our work.

## 3. RESULTS AND DISCUSSION

We characterized MOSFETs and NCFETs with bulk traps based on the model. The device structure of the back-gated 2D AOS NCFETs used in our simulation is schematically shown in Figure 2c. The a-IGZO and HZO were selected as the channel and ferroelectric materials with the thicknesses of $T_{\mathrm{IGZO}} = 5$ nm and $T_{\mathrm{FE}} = 3$ nm, respectively. The pertinent material parameters of IGZO are set as $m_{\mathrm{e}} = 0.255\, m_0$ and $m_{\mathrm{h}} = 3.88\, m_0$, where $m_e$, $m_h$ and $m_0$ are the electron effective, hole effective, and free electron masses, respectively.[37] The constant mobility $\mu$ and band gap were also determined to be 13 $\mathrm{cm^2/Vs}$ and 3.2 eV.[38]

Meanwhile, the thickness of the gate oxide, $SiO_2$, was determined depending on the trap density, considering the following design conditions for NCFETs: the condition for sub-60 $SS$ is $1/|C_{\mathrm{FE}}| > 1/C_{\mathrm{OX}}$, and the condition for hysteresis-free operation is $1/C_{\mathrm{FE}} < 1/C_{\mathrm{OX}} + 1/C_{\mathrm{S},V_{\mathrm{gs}}>V_{\mathrm{th}}}$.[29] Here, the semiconductor capacitance $C_{\mathrm{S}}$ is defined separately for two regions: the bellow-subthreshold region $C_{\mathrm{S},V_{\mathrm{gs}}<V_{\mathrm{th}}}$ and above-threshold region $C_{\mathrm{S},V_{\mathrm{gs}}>V_{\mathrm{th}}}$. This definition arises from the rapid increase in $C_{\mathrm{Q}}$ after charge inversion, which causes $C_{\mathrm{S},V_{\mathrm{gs}}<V_{\mathrm{th}}}$ to expand into $C_{\mathrm{S},V_{\mathrm{gs}}>V_{\mathrm{th}}} = C_{\mathrm{Q}} + C_{\mathrm{S},V_{\mathrm{gs}}<V_{\mathrm{th}}}$. Considering the aforementioned design space of the NCFETs $1/C_{\mathrm{OX}} < 1/C_{\mathrm{FE}} < 1/C_{\mathrm{OX}} + 1/C_{\mathrm{S},V_{\mathrm{gs}}>V_{\mathrm{th}}}$, the minimum $SS$, $SS_{\mathrm{min}} = 60[1 - C_{\mathrm{S},V_{\mathrm{gs}}<V_{\mathrm{th}}}/C_{\mathrm{S},V_{\mathrm{gs}}>V_{\mathrm{th}}}]$, is obtained at the boundary of the hysteresis-free operation, $1/C_{\mathrm{FE}} = 1/C_{\mathrm{OX}} + 1/C_{\mathrm{S},V_{\mathrm{gs}}>V_{\mathrm{th}}}$, which is derived from Eq 10. Under this condition, with a fixed value of $1/|C_{\mathrm{FE}}|$, the addition of $C_{\mathrm{trap}}$ to $C_{\mathrm{S},V_{\mathrm{gs}}>V_{\mathrm{th}}}$ results in a decrease in $1/(C_{\mathrm{S},V_{\mathrm{gs}}>V_{\mathrm{th}}} + C_{\mathrm{trap}})$, which subsequently leads to an increase in $1/C_{\mathrm{OX}}$ (or a decrease in the oxide thickness $T_{\mathrm{OX}}$) to achieve the $SS_{\mathrm{min}}$. Accordingly, to compare each $I_{\mathrm{ds}}$-$V_{\mathrm{gs}}$ characteristic at the $SS_{\mathrm{min}}$, a thinner oxide layer is selected as the trap density increases. To fairly compare NCFETs and MOSFETs, it is necessary to match their gate dielectric properties. When determining the gate oxide layer for NCFETs, we employed the $T_{\mathrm{OX}}$ that yields the $SS_{\mathrm{min}}$ for NCFETs with the given trap density. However, determining the gate dielectric thickness for MOSFETs is not straightforward because the dielectric permittivity of ferroelectric materials varies with the electric field, making the equivalent oxide thickness (EOT) for NCFETs challenging to define. Here, we set the oxide thickness for MOSFETs to the $T_{\mathrm{OX}}$ that yields the $SS_{\min}$ for NCFETs; we would like to note that any larger $T_{\mathrm{OX}}$ simply increases the $SS$ of MOSFETs. This choice allows us to compare the inherent benefit of the NC effect relative to the baseline MOSFET with similar dielectric thickness excluding the FE layer under conditions optimized for the NCFET's steep slope potential at that trap density.

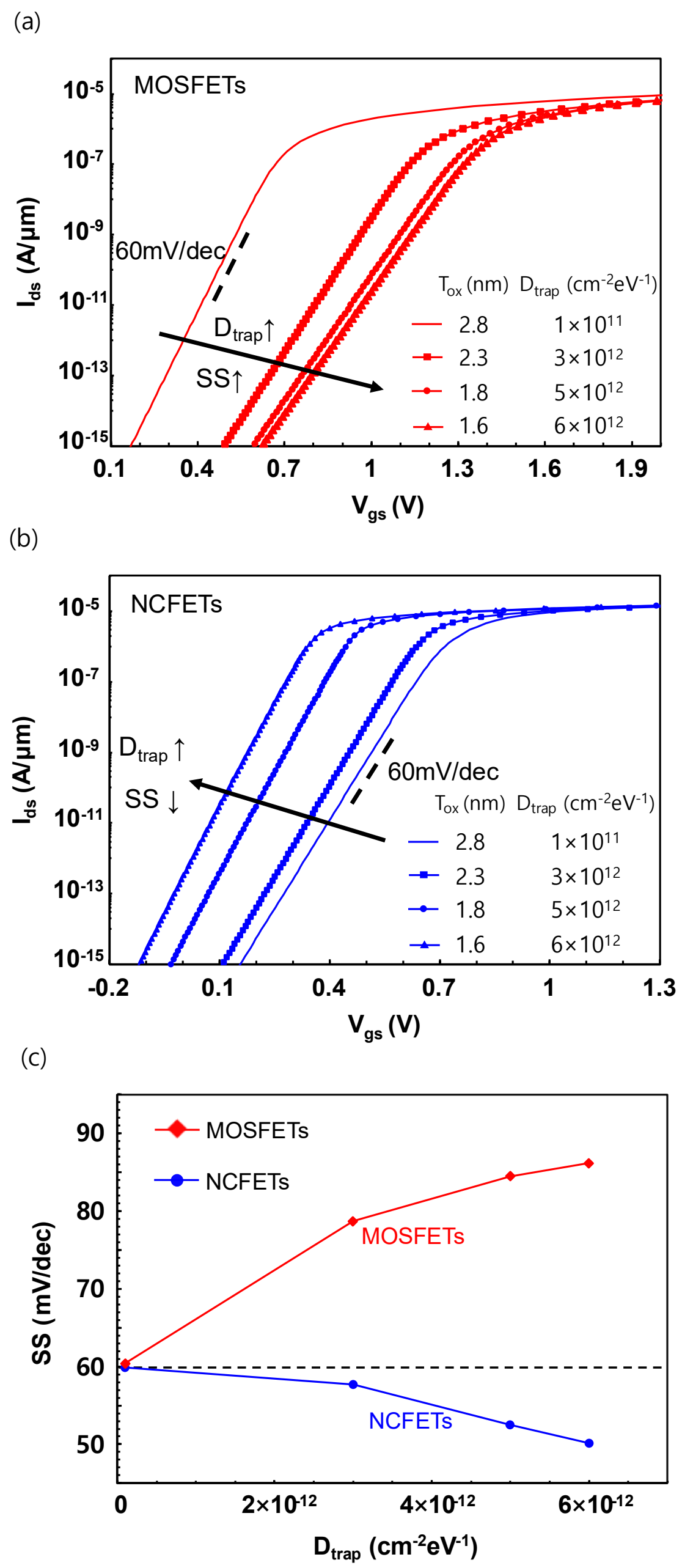

Figure 3. $I_{\mathrm{ds}}$ versus $V_{\mathrm{gs}}$ charaterisitcs of the IGZO (a) NCFETs and (b) MOSFETs for $V_{\mathrm{ds}} = 0.05$ V with different trap densities. The channel length was set to 100 nm, neglecting the short channel effects (SCEs). The channel doping $N_{\mathrm{ch}}$ was determined to be $6.5 \times 10^{7}\ \mathrm{cm}^{-2}$. (c) $SS$ comparison according to the trap densities for both NCFETs and MOSFETs. The dashed line indicates the $SS$ with 60 mV/dec corresponding to the crystalline channel.

Figure 3a shows the $I_{\mathrm{ds}}$-$V_{\mathrm{gs}}$ curves of the MOSFETs according to the $D_{\mathrm{trap}}$, where the $SS$ increases as the $D_{\mathrm{trap}}$ increases. This is because electrons occupying the channel trap states introduce an additional capacitance $C_{\mathrm{trap}}$, leading to the degradation of $SS = 60(1 + C_{\mathrm{trap}}/C_{\mathrm{OX}})$ under the assumption of negligible $C_{\mathrm{S},V_{\mathrm{gs}}<V_{\mathrm{th}}} \approx 0$ in modern FETs. In contrast, Figure 3b shows the improvement in the $SS$ of NCFETs with increasing $D_{\mathrm{trap}}$, while maintaining the conditions of sub-60 $SS$ and hysteresis-free operation. The reduction in $SS$ is attributed to the fact that sufficient $C_{\mathrm{trap}}$ facilitates the full utilization of the NC effect through capacitance matching, as described by $SS = 60[1 - C_{\mathrm{trap}}(1/C_{\mathrm{FE}} - 1/C_{\mathrm{OX}})]$, which is derived from Eq 10. The physical origin is explained by the trapped charges enhancing the negative potential drop of the NC layer, enabling the rapid inversion and the device switching, as discussed afterward.

Figure 3a and b also show the threshold voltage $V_{\mathrm{th}}$ shift in the positive direction for MOSFETs and in the negative direction for NCFETs as the $D_{\mathrm{trap}}$ increases. For the MOSFETs, the addition of $Q_{\mathrm{trap}}$ to $Q_{\mathrm{int}}$ results in an increase in $V_{\mathrm{OX}}(Q_{\mathrm{int}})$, which causes a positive shift in the $V_{\mathrm{th}}$. However, for NCFETs, the sub-60 $SS$ condition provides the relationships $1/|C_{\mathrm{FE}}| > 1/C_{\mathrm{OX}}$, i.e., $|V_{\mathrm{FE}}/Q_{\mathrm{int}}| > V_{\mathrm{OX}}/Q_{\mathrm{int}}$, where the $V_{\mathrm{FE}}$ is negative. Consequently, when $Q_{\mathrm{trap}}$ is added to $Q_{\mathrm{int}}$, $(V_{\mathrm{FE}} + V_{\mathrm{OX}})$ shifts to a negative value, resulting in a corresponding shift of $V_{\mathrm{th}}$ to a negative value as well according to Eq 3.

Figure 3c shows the $SS$ versus $D_{\text{trap}}$ extracted at $I_{\text{ds}} = 1 \times 10^{-11}$ A/μm from Figure 3a and b. In the case of the crystalline channel without traps, the $SS$ is limited to 60 mV/dec under the assumption of the negligible parasitic capacitances. As the trap density increases, the $SS$ in MOSFETs increases, whereas the $SS$ in NCFETs decreases due to the negative $C_{\text{FE}}$. At $D_{\text{trap}} = 6 \times 10^{12}\ \text{cm}^{-2}\text{eV}^{-1}$, the $SS$ in MOSFETs reached $86.2\ \text{mV/dec}$, while the $SS$ in NCFETs achieved $51.5\ \text{mV/dec}$, which is lower than the Boltzmann limit of $60\ \text{mV/dec}$. The incorporation of an NC layer into an AOS transistor can enhance the on/off ratio and potentially mitigate the trap-induced device performance degradation.

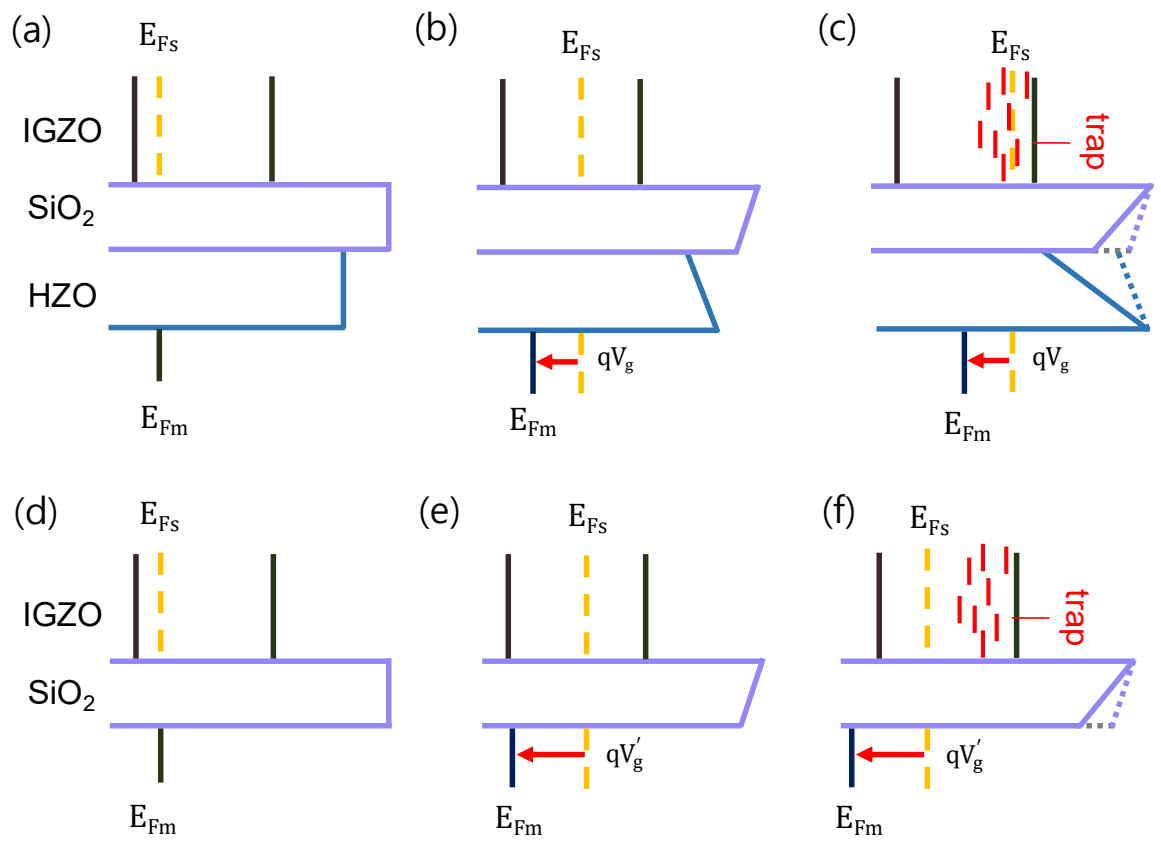

Figure 4. Comparison of the energy band diagrams of (a)-(c) NCFETs and (d)-(f) MOSFETs. (a) and (d) show the band diagram with the flat band voltage. Channel potential shift occurs without traps [(b) and (e)] and with traps [(c) and (f)] under fixed gate bias ($V_g$ for NCFETs and $V_g'$ for MOSFETs). The traps intensify the potential drop in the ferroelectric and $SiO_2$ layers and induce a band shift in the IGZO channel of the NCFETs and MOSFETs in opposite directions.

The underlying mechanism of the steep slope operation resulting from the traps can be elucidated as follows. In NCFETs, the negative voltage drop of $V_{\text{FE}}$ compensates for the positive voltage drop of the internal voltage $V_{\text{int}} = V_{OX} + \psi_s$, which enables sub-60 $SS$. The $SS$ can be expressed as[29]:

$$SS = 60\left[1 - C_{\mathrm{S},V_{\mathrm{gs}}<\mathrm{V_{th}}}\left(\frac{1}{|C_{\mathrm{FE}}|} - \frac{1}{C_{\mathrm{OX}}}\right)\right]. \tag{10}$$

The series connection of $|C_{\mathrm{FE}}|$ and $C_{\mathrm{OX}}$, as shown in the capacitor divider circuit in Figure 2a, determines $SS$, which can be reduced to less than 60 mV/dec when $1/|C_{\mathrm{FE}}| > 1/C_{\mathrm{OX}}$. It should be noted that $C_{\mathrm{S},V_{\mathrm{gs}}<V_{\mathrm{th}}}$, including parasitic capacitances, such as the depletion capacitance and the gate-to-source capacitance, is required to be minimized for the modern FET design, and the minimized $C_{\mathrm{S},V_{\mathrm{gs}}<V_{\mathrm{th}}}$ makes it difficult to achieve a low $SS$ for NCFETs. Therefore, the $SS$ is limited to 60 mV/dec in both NCFETs and MOSFETs with an ultrathin crystalline channel. However, if a significant amount of traps exist in the AOS channel, the $SS$ in NCFETs can be improved due to the trap capacitance $C_{\mathrm{trap}}$ connected in parallel to $C_{\mathrm{S},V_{\mathrm{gs}}<V_{\mathrm{th}}}$, as illustrated in Figure 2a. Conversely, the $SS$ in MOSFETs is degraded because $C_{\mathrm{trap}}/C_{\mathrm{OX}}$ >1.

To provide further physical interpretation regarding sub-60 $SS$ by traps in NCFETs, a comparison was conducted by analyzing the band shift of NCFETs and MOSFETs, depending on the presence or absence of $Q_{\mathrm{trap}}$ under the fixed gate bias as shown in Figure 4. The voltage division in Eq 3 can be rewritten as a function of charges as follows:

$$V_{\mathrm{gs}} = V_{\mathrm{fb}} + V_{\mathrm{FE}}(Q_{\mathrm{int}}) + V_{\mathrm{OX}}(Q_{\mathrm{int}}) + \psi_{\mathrm{s}}(Q_{\mathrm{inv}}). \tag{11}$$

The added $Q_{\mathrm{trap}}$ enhances the polarization response $|P|$ for the given $Q_{\mathrm{int}}$, leading to the large magnitude negative shift in $V_{\mathrm{FE}}$. Since $V_{\mathrm{gs}} = V_{\mathrm{fb}} + V_{\mathrm{FE}} + V_{\mathrm{OX}} + \psi_{\mathrm{s}}$ is fixed, and $V_{\mathrm{OX}}$ changes relatively less than $V_{\mathrm{FE}}$ (due to $1/|C_{\mathrm{FE}}| > 1/C_{\mathrm{OX}}$), the more negative $V_{\mathrm{FE}}$ necessitates the more positive $\psi_{\mathrm{s}}$ to maintain the voltage valance, resulting in stronger inversion for the same $V_{\mathrm{gs}}$. Finally, the large band shift, together with the $\psi_{\mathrm{s}}$, allows us to achieve rapid inversion, enabling early device onset and the steep slope. By contrast, for MOSFETs, the $Q_{\mathrm{trap}}$ induces an additional positive shift in $V_{\mathrm{OX}}$, necessitating a large positive shift in $\psi_{\mathrm{s}}$, i.e., more band shift, to achieve inversion. Therefore, higher gate bias is required for switching the device, resulting in the $SS$ degradation.

## 4. CONCLUSION

We presented a compact 2D NCFET model incorporating the trap states based on the L-K equation and the charge distribution dependence along the channel direction. Then, we investigated the transport characteristics with different trap densities. The simulation results demonstrated that the bulk traps in the NCFETs result in the $SS$ improvement and the $V_{\mathrm{th}}$ shift in a negative direction, in contrast to the behavior observed in the MOSFETs. In addition, we quantitatively analyzed the contribution of traps in steep-slope operations by presenting the band diagram depending on the existence of the traps. Our analysis provides insight into engineering the AOS NCFETs with a steep-slope.

Although a clear interpretation of the $SS$ improvement has been provided, further discussion on the various types of traps is needed, as the current work only considers uniform bulk traps. In fact, various trap distributions such as Gaussian and exponential types can exist in IGZO; however, if the trap distribution is energetically broad, comparable NC effects and corresponding SS improvement can be expected. Moreover, defects may exist not only in the oxide bulk but also the ferroelectric-dielectric interface.[23] Extrinsic defects introduced by doping and intrinsic defects within the gate stack can also act as trap states. Thus, further investigation into these other trap states would be beneficial for the steepness.

In our model, simulations were conducted using a long channel length to neglect SCE. However, short-channel devices exhibit a significantly higher source/drain geometrical capacitance compared to long-channel FETs with minimal parasitic capacitance.[29] The increase in $C_{\mathrm{S}}$ leads to the $SS$ reduction. Given the continued scaling of semiconductor devices, developing a model that integrates both channel length effects and trap effects would be a valuable contribution to the field.

Although our work focused solely on bulk traps without considering other trap types or geometrical effects, it is noteworthy that we achieved an improvement in $SS$ through the use of traps, overturning the conventional perception that traps degrade device performance. Our findings are not confined to the AOS applications but can be extended to broader research on device applications that employ channel materials with traps, including various amorphous materials and polycrystalline structures. This insight into trap-enhanced NC behavior opens up possibilities for the practical application of trap-rich materials such as polycrystalline silicon and amorphous semiconductors, including display backplanes, TFTs, and flexible electronics. Ultimately, this work will contribute to the optimization of the AOS device performance through the NC effect induced by ferroelectric polarization and will also accelerate the advanced semiconductor technology, such as M3D integration, 3D NAND, and BEOL-compatible devices.

AUTHOR INFORMATION

**Corresponding Author**

*Corresponding author. E-mail: y.lee@inu.ac.kr

**Author Contributions**

The authors are listed in order of contribution to the work.

All authors have given approval to the final version of the manuscript.

**ACKNOWLEDGMENT**

This work was supported by the Incheon National University Research Grant in 2023.

The EDA tool used for validating our model was supported by the IC Design Education Center.

(Corresponding author: Yeonghun Lee.)

**Notes**

The authors declare the following competing financial interest(s): The authors have filed a patent application with the Korean Intellectual Property Office based on the findings presented in this article.

## TOC

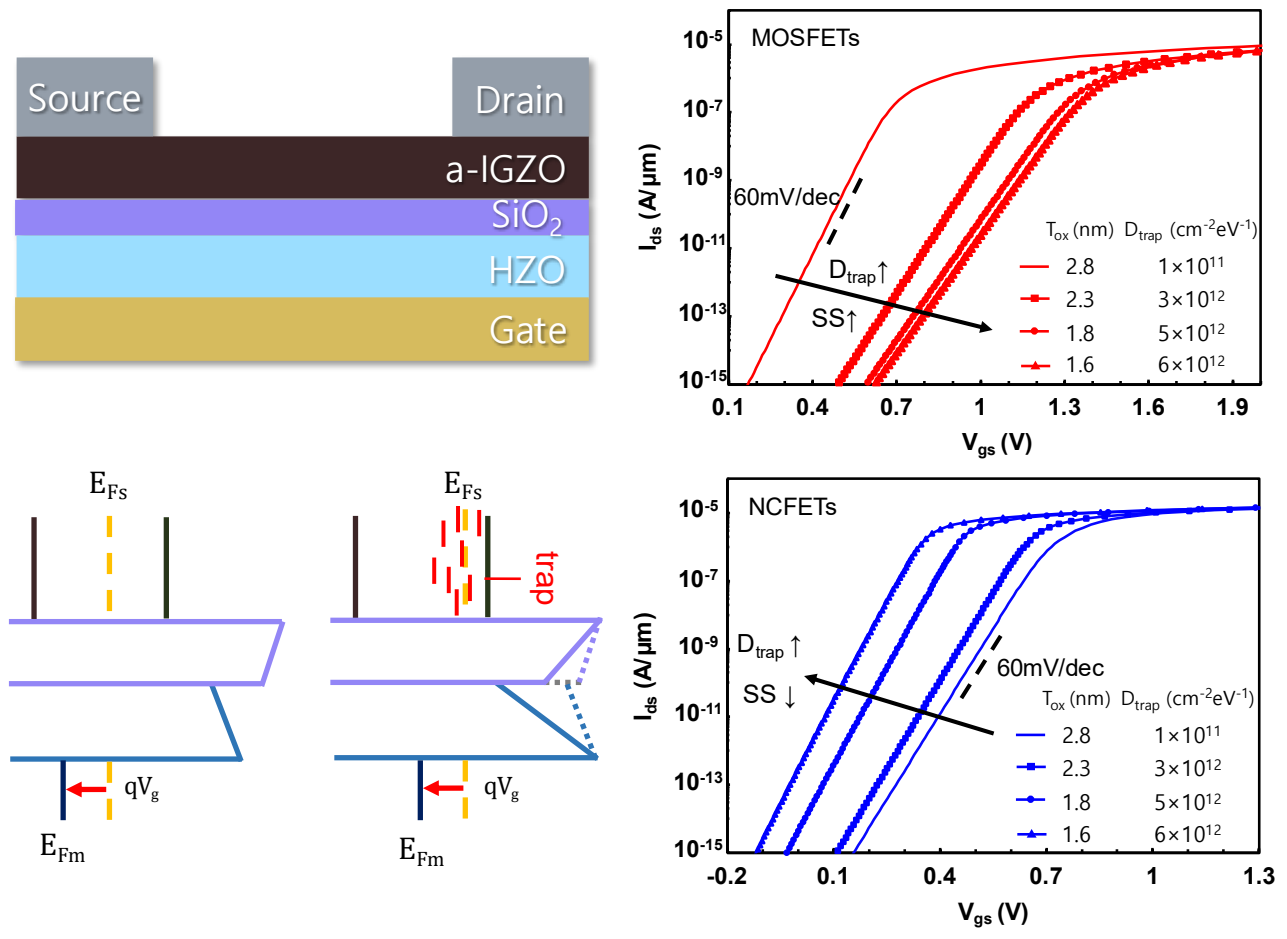